\documentclass[aip,amsmath,amssymb,reprint]{revtex4-1}
\usepackage{epsfig}
\usepackage{amsmath}
\graphicspath{{figures/}}
\usepackage{epsfig}
\usepackage{amsmath}
\usepackage{color}
\usepackage{graphicx,epsfig}
\usepackage{color}
\usepackage[latin9]{inputenc}
\usepackage{float}
\usepackage{natbib}
\begin{document}

\title{Melting points of water models: current situation }

\author{S. Blazquez}
\author{C. Vega$^{*}$}
\affiliation{$^1$Dpto. Qu\'{\i}mica F\'{\i}sica I, Fac. Ciencias Qu\'{\i}micas,
Universidad Complutense de Madrid, 28040 Madrid, Spain.}

%\date{\today}
%
\begin{abstract}

By using the direct coexistence method  we have calculated the melting points of 
Ice Ih at normal pressure for three recently proposed water models, namely 
TIP3P-FB, TIP4P-FB and TIP4P-D.  We obtained 
T$_m$=216 K for TIP3P-FB, T$_m$=242 K for TIP4P-FB and T$_m$=247 K for TIP4P-D. We revisited 
the melting point of TIP4P/2005 and TIP5P obtaining T$_m$=250 and 274 K respectively. 
We summarize the current situation of the melting point of ice Ih for a number 
of water models and conclude that no model is yet able to  simultaneously reproduce  
the melting temperature of ice Ih and the temperature of the maximum in density at room 
pressure. This probably points towards our both still incomplete knowledge of the potential 
energy surface of water and the necessity of incorporating nuclear quantum 
effects to describe both properties simultaneously. 
\end{abstract}

\maketitle
$^*$Corresponding author: cvega@quim.ucm.es

When simulating water a force field is needed and it is common to use
a simple rigid non-polarizable force field. Usually positive charges are located on the hydrogen atoms
and a Lennard-Jones center is located on the oxygen.  In three center models (3C) the negative 
charge is located on the oxygen atom as in the TIP3P\cite{jorgensen83}, SPC\cite{spc} and SPC/E\cite{berendsen87},
models proposed by the groups of Jorgensen and Berendsen respectively. 
In the four center models (4C) the 
negative charge is located along the bisector of the H-O-H angle\cite{bernal33}
leading to the popular TIP4P model\cite{jorgensen83}. 
In this first wave of water models (80's) the density and vaporization enthalpy were used as target properties (although
in the case of SPC/E only when including the self energy correction\cite{berendsen87}).
Water, at constant pressure, has a temperature at which the density reachs a maximum (TMD). 
Recognizing the importance of that led to the second wave (2000-2010) of potential models where the TMD was 
used as a target property.  Two different approaches to achieve this goal 
were used. In the first approach  
the negative charge was located in the position of the lone pair electrons as in the TIP5P\cite{mahoney01} 
(a geommetry also used in the old ST2 model\cite{stillinger74}) 
resulting in five center models (5C). In the second approach the TIP4P geometry was kept but the vaporization enthalpy
was  sacrificed as a target property (unless the self energy correction is included)  
in favour of the TMD as in the TIP4P-Ew \cite{tip4p-ew} and TIP4P/2005 models\cite{abascal05b}. 
Over the last ten years some additional non polarizable models have been proposed. Some of then 
using a 3C geometry as  OPC-3\cite{izadi2016accuracy} or TIP3P-FB\cite{doi:10.1021/jz500737m} and some of them using a 4C geometry as 
TIP4P-FB\cite{doi:10.1021/jz500737m}, TIP4P-$\epsilon$\cite{doi:10.1021/jp410865y},TIP4P-D\cite{doi:10.1021/jp508971m} or 
OPC\cite{izadi14}. In general the aim of these models was  to improve the description 
of the dielectric constant of liquid water (but not in ices) with respect to TIP4P-Ew and TIP4P/2005 models although in general 
the improvement was made at the cost of deteriorating the predictions for another property (see Ref.\cite{vegamp15} 
for a general discussion on the role of the dielectric constant in water simulations). 
If we focus on polarizable models we can also find three new and interesting force fields
such as the BK3\cite{kiss13}, i-AMOEBA\cite{doi:10.1021/jp403802c}, MB-Pol\cite{mbpol_paesani} or HBP\cite{doi:10.1021/acs.jpcb.6b08205}
models which 
add polarization to a 4C geometry.
  Not only the TMD (at 1 bar) is an interesting property of water but also the melting point of ice Ih (also at 1 bar). 
In fact one of the properties which one can study to validate 
a water force field is the melting point of ice I$_h$.
For most of the models proposed up to 2012 the melting point of ice is well known.\cite{JCP_2006_124_144506,vega05a}
For some of the models proposed over the last ten years we know now the melting point, as for instance 
for OPC and OPC-3 \cite{doi:10.1021/acsomega.0c02638} and for TIP4P-$\epsilon$\cite{doi:10.1021/jp410865y}.
However the melting point of ice Ih for three popular recently proposed water models, namely 
TIP3P-FB\cite{doi:10.1021/jz500737m}, TIP4P-FB\cite{doi:10.1021/jz500737m} and TIP4P-D\cite{doi:10.1021/jp508971m}
is unknown. The goal of this work is to determine their melting temperatures
and to summarize the current situation of force fields with respect to their capacity to predict 
the melting point of ice and the TMD. 
We will use the direct coexistence method\cite{JCP_2006_124_144506,C1CP21210A}, where a solid 
phase consisting of 2000 molecules of ice Ih (proton disordered configuration was 
obtained using the algorithm of Buch et al.\cite{buch98})  is placed in contact with 2000 molecules of liquid water. 
The ice plane exposed at the interface is the secondary prismatic one (1$\bar{2}$10).
We have performed anisotropic $NpT$ simulations with GROMACS package\cite{spoel05} with a time step of 2 fs.
%Periodic boundary conditions in all directions were also applied  in all runs.
Temperature and pressure were kept constant by using the 
 Nos\'e-Hoover thermostat\cite{hoover85}
and Parrinello-Rahman barostat\cite{parrinello81} both with a coupling constant of 2 ps.
%The three
%different sides of the simulation box were allowed to fluctuate independently to allow changes in the shape of the solid
%region and to avoid the existence of stress in the solid.
For electrostatics and Van der Waals interactions the cut-off radii was fixed at 1.0 nm and long-range
corrections to the LJ part of the potential in the energy and pressure were applied. We used PME \cite{essmann95} to account for
the long-range electrostatic forces and LINCS\cite{hess08b} for constraints.

\begin{center}
\begin{figure}[!hbt] \centering
    \centering
    \includegraphics*[clip,scale=0.35,angle=0.0]{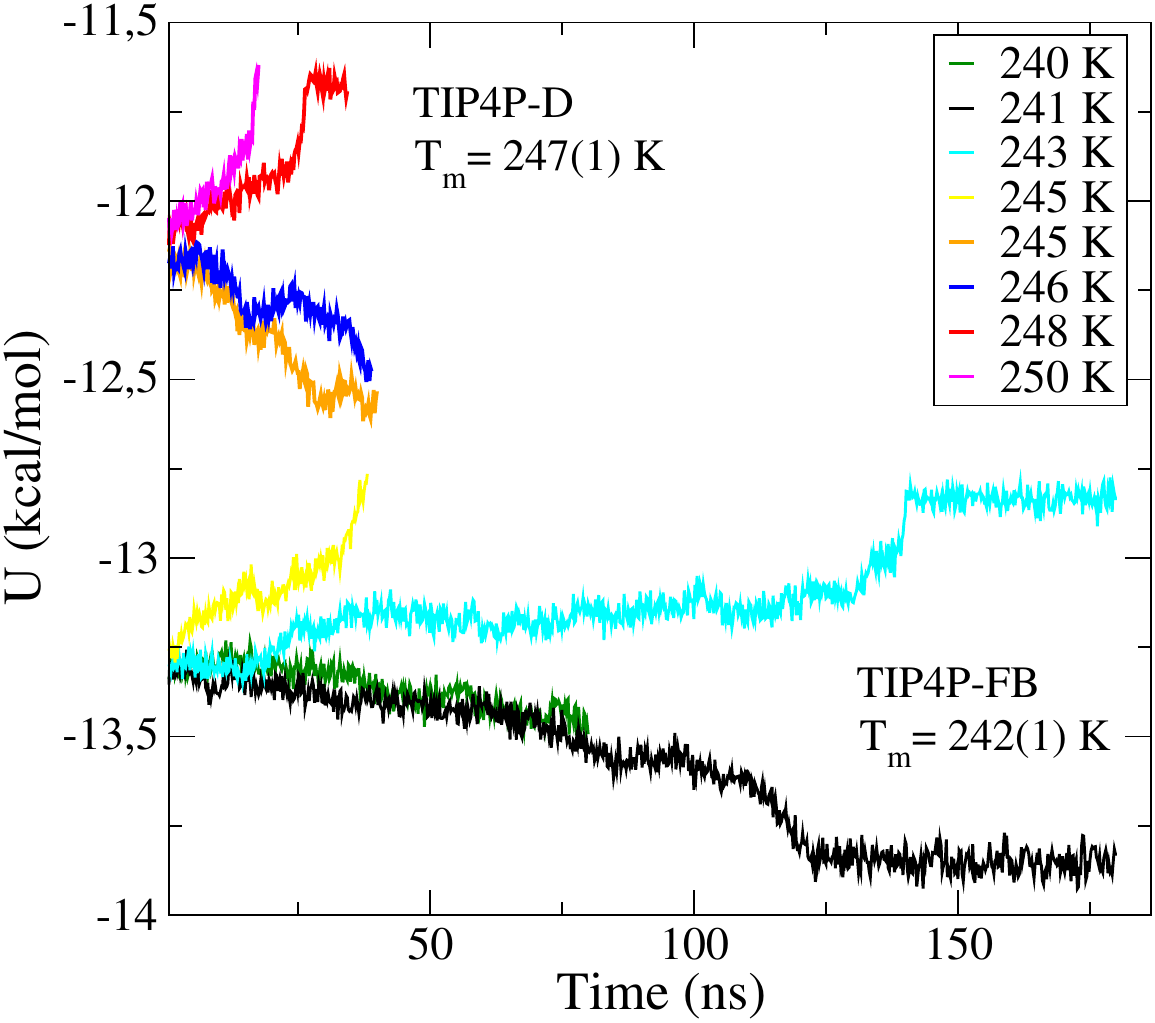}
        \caption{ Evolution of the potential energy as a function of time for the NpT
	runs of TIP4P-D (top) and TIP4P-FB (bottom) models at 1 bar and different temperatures.
	The energies of TIP4P-D water model are shifted 1.3 kcal/mol for better visualization of the reader.}
    \label{energies-ices}
\end{figure}
\end{center}

In Figure \ref{energies-ices} the time evolution of the potential energy of
the system at several temperatures for  TIP4P-FB and TIP4P-D force fields is shown.
For the TIP4P-FB model
ice melts (i.e energy grows) for all temperatures above 243 K. Ice grows at all temperatures
below 241 K. Thus, we can conclude that
the melting temperature for the TIP4P-FB model is T$_m$ = 242(1) K (not surprisingly
similar to that of  TIP4P/$\epsilon$  T$_m$ = 240(1) K taking into account the similarity of the parameters
of both models).
Following the same procedure, we estimate that the melting temperature
of TIP4P-D force field is T$_m$ = 247(1) K. To evaluate the impact of the
cutoff on the calculations we repeated for TIP4P-D the calculations using a larger cutoff (i.e 1.2 nm) obtaining
again  T$_m$ = 247(1) K.

\begin{center}
\begin{figure}[!hbt] \centering
    \centering
    \includegraphics*[clip,scale=0.3,angle=0.0]{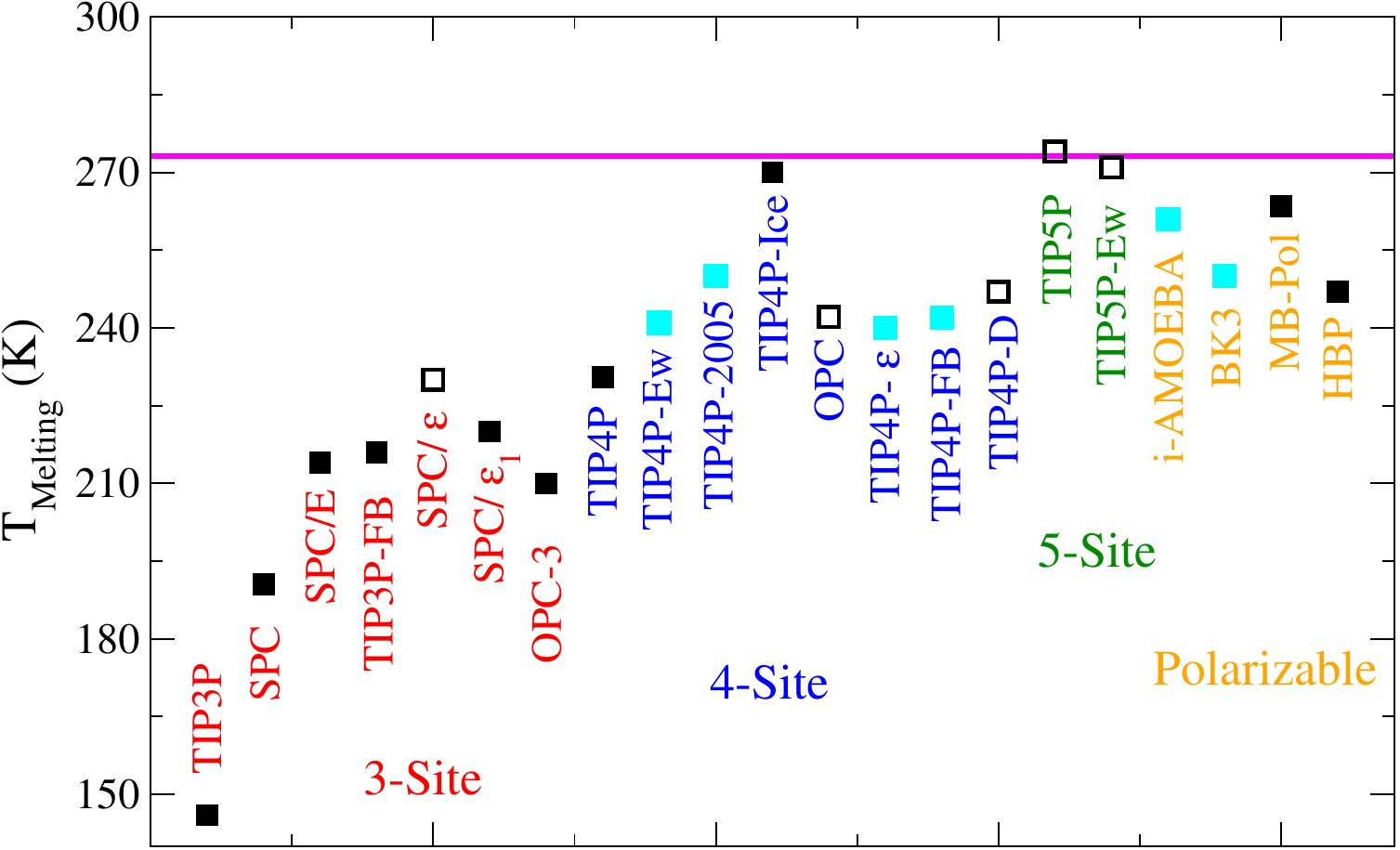}
	\caption{ Melting points of ice I$_h$ of different water models at 1 bar. Blue filled squares: Models 
	which provide (at 1 bar) a good estimation of the TMD (maximum of 4 K of deviation from the experiment).
	Empty squares: Models which provide a 
	fair estimation of the TMD (maximum of 5-10 K of deviation from the experiment). 
	Black filled squares: Models with a bad estimation of the TMD (more than 10 K deviation from the 
	experiment).}
    \label{resumen-tm}
\end{figure}
\end{center}

%As can be seen in  \ref{energies-ices} the growth rate of ice 
%is faster in  TIP4P-D  than in TIP4P-FB (being the growth rate of ice for TIP4P/2005 intermediate
%between these two models as shown in the Supplementary Material). 
We have also recalculated the melting point of TIP4P/2005 and TIP5P 
using the same system size obtaining 
250(1) and 274(1) K respectively (see Supp. Mat.). The melting point
of TIP4P/2005 is in 
excellent agreement with the result reported by Conde et al.\cite{doi:10.1063/1.5008478}  
Finally, we also evaluated the melting point of the TIP3P-FB obtaining (see supplementary material) 
T$_m$ = 216(4) K (the larger error bar is due to the slow dynamics at such low temperatures). 
We have also determined the melting enthalpy at the melting temperature for  
TIP3P-FB, TIP4P-FB, OPC, TIP4P-D and TIP5P obtaining 0.63, 0.99, 1.07, 1.11 and 1.78 kcal/mol respectively, compared 
with the result obtained for TIP4P-2005 (i.e 1.13 kcal/mol) and the experimental value (i.e 1.44 kcal/mol). 
Let us now present a more general discussion. In Figure  \ref{resumen-tm} the results of the melting 
point of water models are  presented. 
In Table \ref{t-melting-water} we also show the numerical results for the melting points,
the TMD of the models and the difference in temperature between the TMD and the melting temperature ($\Delta$T = TMD - T$_m$).
As can be seen in Fig. \ref{resumen-tm}, 3C models yield a poor description of the melting temperature of ice Ih (the 
average being located around 220 K). In short, 3C are not recommended to study the freezing of water (in addition 
ice Ih may not be the most stable phase at room pressure for these models\cite{vega09}). 4C models improve the description, the average melting temperature 
being around 245 K. 
As can be seen the melting points of TIP4P-FB and TIP4P-D are below that of TIP4P/2005. 
Polarizable models using a TIP4P geometry improve the description of the melting point, the average being located around 
255 K but still below the experimental value. The only models reproducing the experimental value are those of 
the 5C geometry, the coarse grained mW\cite{doi:10.1021/jp805227c}, the TIP6P\cite{doi:10.1063/1.4973000} and the special purpose model TIP4P/Ice.\cite{JCP_2005_122_234511}  
In general the melting point  increases with the value of the quadrupole moment of the model.\cite{abascal07c}

It is interesting to analyze the performance of the models with respect to the TMD. In Fig. \ref{resumen-tm}, 
models with a small deviation (4 K or less from the experimental value) are represented as blue squares, 
models with a moderate deviation (i.e between 5 and 10 K) are represented by  empty squares and 
models with a large deviation from experiment (more than 10 K) are represented by black squares. 
As can be seen models that reproduce well the melting point do not reproduce well the TMD and viceversa. 
For most of the models the difference in temperature between the TMD and the melting temperature is too large, ranging 
from 11 to 50 K when compared to the experimental value which is only 4 degrees 
(the only exceptions are the MB-Pol and HBP models for which this difference is almost 0 and 7 K respectively).

\begin{table}[!hbt]
\begin{center}
	\caption{\label{t-melting-water} Melting temperature of ice I$_h$ (T$_m$) and temperature of the maximum in density (TMD) both at 1 bar 
	for different water models as calculated in this work or taken from the literature.
We also show the difference between the TMD and the T$_m$ ($\Delta$T). TMD uncertainty is tipically 2 K.}
\begin{tabular}{ l c c c }
\hline
\hline
	Model & T$_m$ (K) & TMD (K) & $\Delta$T (K)\\
\hline
	\textbf{Expt. }  &273 & 277 & 4\\
	\textbf{TIP3P }  &146(5)\cite{vega05a} & 182\cite{vega05b} & 36\\
	\textbf{ SPC }   &190.5(5)\cite{vega05a} &228\cite{vega05b} & 37.5\\
	\textbf{ SPC/E }   &214(3)\cite{JCP_2006_124_144506} &241\cite{vega05b} & 27 \\
	\textbf{TIP3P-FB } &216(4) This work &261\cite{doi:10.1021/jz500737m} & 45\\
	\textbf{SPC-$\epsilon$ }  &230(2)\cite{FUENTESAZCATL2015116} & 270\cite{FUENTESAZCATL2015116} & 40\\
	\textbf{SPC-$\epsilon_2$ }  &220(2)\cite{FUENTESAZCATL2015116} & 250\cite{FUENTESAZCATL2015116} & 30\\
	 \textbf{OPC-3 }  &210(10)\cite{doi:10.1021/acsomega.0c02638} &260\cite{izadi2016accuracy} & 50 \\
	\textbf{TIP4P }  &229(9)\cite{doi:10.1063/1.1801272} &253\cite{vega05b} & 24\\
        \textbf{TIP4P-FQ }  &303(8)\cite{NICHOLSON200678} &253\cite{rick01} & -50\\
	\textbf{TIP4P-Ew }  &241(1)\cite{doi:10.1021/acsomega.0c02638} &274\cite{tip4p-ew} & 33\\
	\textbf{TIP4P-2005 }  &250(1) This work &277 \cite{pi09} & 27\\
	\textbf{TIP4P-Ice }  &270(3)\cite{JCP_2006_124_144506} & 295\cite{vega05b} & 25\\
	\textbf{OPC }  &242.2(0.9)\cite{doi:10.1021/acsomega.0c02638}&272\cite{izadi14} & 29.8 \\
	\textbf{TIP4P-$\epsilon$ }  &240(2)\cite{doi:10.1021/jp410865y} & 276\cite{doi:10.1021/jp410865y} & 36\\
	\textbf{TIP4P-FB }  &242(1) This work &277\cite{doi:10.1021/jz500737m} & 35\\
	\textbf{TIP4P-D }  &247(1) This work &270\cite{doi:10.1021/jp508971m} & 23\\
	\textbf{TIP5P}  &274(1) This work &285 \cite{lisal02} & 11 \\
	\textbf{TIP5P-EW}  &271(3)\cite{JCP_2006_124_144506}  &282 \cite{rick04} & 11\\
	\textbf{ST2}  &299(2)\cite{doi:10.1080/00268976.2015.1043966}  &323 \cite{poole05} & 24\\
	\textbf{TIP6P}  &274.5(1.5)\cite{doi:10.1063/1.4973000}  &290 \cite{doi:10.1063/1.4973000} & 15.5\\
	\textbf{mW}  &273(1.5)\cite{mw-properties}  &251 \cite{mw-properties} & -22\\
%        \textbf{AMOEBA}  & 292\cite{doi:10.1021/jp0484332} \\
	\textbf{i-AMOEBA}  & 261(2)\cite{doi:10.1021/jp403802c} & 277\cite{doi:10.1021/jp403802c} &16 \\
%        \textbf{BKd1}  & 251\cite{doi:10.1063/1.4746419}\\
%        \textbf{BKd2}  & 237\cite{doi:10.1063/1.4746419} \\
%        \textbf{BKd3}  & 278\cite{doi:10.1063/1.4746419} \\
	\textbf{BK3}  & 250(3)\cite{kiss13} & 275\cite{kiss13} & 25\\
	\textbf{MB-Pol}  & 263.5(1.5)\cite{mbpol_paesani} & 263\cite{doi:10.1021/acs.jpclett.2c00567} & -0.5 \\
        \textbf{HBP}  & 247(3)\cite{doi:10.1021/acs.jpcb.6b08205} & 254\cite{doi:10.1021/acs.jpcb.6b08205}  & 7 \\
\hline
\hline
    \end{tabular}
  \end{center}
\end{table}

To summarize, in this work we have determined  the melting points of
three recently proposed rigid and non polarisable water models (TIP4P-FB, TIP4P-D and TIP3P-FB) with the goal 
of analyzing if they could reproduce simultaneously the melting point and the TMD. 
The answer is negative and their melting points are similar (although a few degrees lower) than that of TIP4P/2005. 
The main conclusion is that in 2022 we do not have yet any model of water 
(polarizable or not) able to reproduce  simultaneously both the melting point of ice Ih and the TMD. 
Further work is needed to determine whether this is a deficiency in the description of the potential energy
surface of water (PES) of all force fields proposed so far, or to the necessity of incorporating nuclear quantum effects to 
describe both properties at the same time accurately.\cite{nuclear_quantum_effects,doi:10.1021/acs.chemrev.5b00674,guillot02}
If this were the case then even an accurate PES could not reproduce both 
properties simultaneously when using classical simulations. 

See the Supplementary Material for aditional figures of the melting
point of the models studied in this work and for the comparison of different properties
such as melting enthalpy and densities of ice and liquid water for each model.

%\noindent Funding from grant PID2019-105898GB-C21 is acknowledged
This project has been funded by grant PID2019-105898GB-C21
of the  Ministerio de Educacion y Cultura.

\section{Conflict of Interest}
The authors have no conflicts to disclose.

\section{Data Availability}
The data that support the findings of this study are available
within the article and in the Supp. Mat.

\end{document}